\documentclass[conference]{IEEEtran}
  \usepackage{cite}
  \usepackage{amsmath}
  \usepackage{amssymb}
  \usepackage{latexsym}
  \usepackage{epsfig}
  \usepackage{graphics}
  \usepackage{color}

\ifCLASSINFOpdf

\else

\fi

\newcommand\bb{\boldsymbol{b}}

\newcommand\yb{\boldsymbol{y}}
\newcommand\hb{\boldsymbol{h}}

\newcommand\nb{\boldsymbol{n}}

\newcommand\ub{\boldsymbol{u}}
\newcommand\vb{\boldsymbol{v}}

\newcommand\Bb{\mathbb{B}}

\newcommand\Db{\mathbb{D}}

\newcommand\Ib{\mathbb{I}}

\newcommand\Qb{\mathbb{Q}}

\newcommand\Xb{\mathbb{X}}

\newcommand\tr{\tt{tr}}
\renewcommand\det{\tt{det}}

\newcommand\TT{{\tt T}}
\newcommand\HH{{\tt H}}

\definecolor{cyan}{rgb}{0, 1, 1}
\definecolor{green}{rgb}{0, 1, 0}
\definecolor{red}{rgb}{1, 0, 0}
\definecolor{blue}{rgb}{0, 0, 1}
\definecolor{purple}{rgb}{0.5, 0, 0.5}
\definecolor{darkred}{rgb}{0.565, 0.0, 0.0}
\definecolor{lightgray}{rgb}{0.8, 0.8, 0.8}
\definecolor{lightred}{rgb}{1, 0.5, 0.5}
\definecolor{lightblue}{rgb}{0.5, 0.5, 1}

\newcounter{step}

\begin{document}

\title{A Two-Phase Maximum-Likelihood Sequence Estimation for Receivers with Partial CSI}

\author{
\authorblockN{Chia-Lung Wu\authorrefmark{1}, Po-Ning~Chen\authorrefmark{2},
Mikael Skoglund\authorrefmark{1}, Ming Xiao\authorrefmark{1} and
Shin-Lin Shieh\authorrefmark{3}}
\authorblockA{\authorrefmark{1}School of Electrical Eng., Royal Institute of Technology, Stockholm, Sweden \\
Email: clw@kth.se, skoglund@ee.kth.se, mingx@kth.se}
\authorblockA{\authorrefmark{2}Dept.~of Electrical and Computer Eng., National Chiao-Tung Univ., Taiwan, ROC\\
Email: poning@faculty.nctu.edu.tw}
\authorblockA{\authorrefmark{3}Graduate Institute of Comm.~Eng., National Taipei Univ., Taiwan, ROC\\
Email: slshieh@mail.ntpu.edu.tw}}

\maketitle

\begin{abstract}
The optimality of the conventional maximum-likelihood sequence
estimation (MLSE), also known as the Viterbi Algorithm (VA),
relies on the assumption that the receiver has perfect knowledge
of the channel coefficients or channel state information (CSI).
However, in practical situations that fail the assumption,
the MLSE method becomes suboptimal and then
exhaustive checking is the only way to obtain the ML sequence. At
this background, considering directly the ML criterion for
partial CSI, we propose a two-phase
low-complexity MLSE algorithm, in which the first
phase performs the conventional MLSE algorithm in order to retain
necessary information for the backward VA performed in the second
phase. Simulations show that when the training sequence is
moderately long in comparison with the entire data block such as
1/3 of the block, the proposed two-phase MLSE can approach the
performance of the optimal exhaustive checking. In a normal case,
where the training sequence consumes only 0.14 of the bandwidth,
our proposed method still outperforms evidently the conventional
MLSE.
\end{abstract}

\bigskip

\section{Introduction}

In order to combat the signal distortion due to inter-symbol
interference in frequency-selective fading channels, a receiver
generally needs a channel estimator and an equalizer, where the
former estimates the channel state information (CSI) based on a
training sequence, while the latter performs the detection of data
using the CSI obtained by the former. In the literature, a
commonly used equalization method is the Euclidean-distance-based maximum-likelihood
sequence estimation (MLSE) \cite{Forney}. This MLSE is optimal if
the estimator can perform perfect channel estimation; however,
when the channel estimator cannot pass perfect CSI to the
equalizer, the system performance degrades, thereby inducing the
research about receivers with only partial CSI.

The detection criterion for a receiver with only partial CSI,
usually referred to as  \emph{partially coherent} receiver, has
been investigated in \cite{CWS,Viterbi,Giese,Zhang0,Coskun}.
Specifically, they found that the ML criterion for a partial
coherent receiver can actually be written as a weighted sum of the
ML criterion assuming perfect CSI in the receiver and the ML
criterion that assumes no CSI available in the receiver. Since
exhaustive checking is the unique optimal method for performing ML
sequence estimation for a receiver without CSI, their finding
makes the usual Viterbi algorithm (VA) unsuitable for optimal
sequence detection when only partial CSI is available
\cite{Coskun}.

For this reason, we propose in this paper a two-phase method
to perform the sequence estimation for a partially coherent receiver.
 In short,  the forward VA will be executed in the first phase, generating
 necessary information required by the backward VA that
 uses the partial-CSI ML criterion
 in the second phase. Simulation results confirm that the proposed two-phase
 method can considerably outperform the conventional MLSE over channels with only partial CSI available.

Throughout this paper, the following notations will be used: For a
matrix $\Xb$, $\det|\Xb|$ is its determinant; $\Xb^\TT$ and
$\Xb^\HH$ denote its transpose and Hermitian transpose,
respectively. Also, $\Ib$ will be used to denote the identity
matrix of a proper size.

\medskip

\section{System Model}
\label{systemmodel}

In this paper, we consider a signal $\bb=[b_1,\ldots,b_N]^\TT$
transmitted over a frequency-selective block fading (equivalently,
{\em quasi-static} fading) channel of memory order $P-1$. For
$1\leq i \leq N$ and $M>0$, we restrict that $b_i$ is the output
of constant-amplitude $(2^M)$-PSK modulation, and hence
$|b_i|^2=1$. Among the $N$ components in signal $\bb$, the first
$T$ components are the training sequence and are assumed known to
the receiver, while the latter $(N-T)$ symbols are the data to be
transmitted. The received vector $\yb$ can thus be
\begin{eqnarray}
\yb=\Bb\hb+\nb,\label{channel_model}
\end{eqnarray}
where
$$\Bb=\begin{bmatrix}\Bb_{P} \\
\Bb_D\end{bmatrix}$$
is formed by a $(T \times P)$ submatrix
$\Bb_P$ and a $((L-T) \times P)$ submatrix $\Bb_D$,
which are respectively defined as
$$\Bb_P\triangleq\begin{bmatrix}
    b_1        & \cdots & 0\\
    b_2        & \ddots & \vdots\\
    b_3        & \ddots & 0\\
    b_4        & \ddots & b_1\\
    \vdots     & \ddots & \vdots\\
    b_T        & \cdots & b_{T-P+1}
  \end{bmatrix},$$ and
  $$
\Bb_D\triangleq\begin{bmatrix}
    b_{T+1}& \cdots & b_{T-P+2}\\
    \vdots & \ddots & b_{T-P+3}\\
    b_N    & \ddots & \vdots\\
    0      & \ddots & b_{T}\\
    \vdots & \ddots & \vdots\\
    0      & \cdots & b_N
  \end{bmatrix}.$$
In \eqref{channel_model}, noise $\nb$ is zero-mean circular symmetric complex Gaussian distributed
with correlation matrix $\sigma^2_n\Ib$, and
$\hb=[h_1,\ldots,h_P]^T$ denotes the channel taps that remain constant
during an $L$-symbol transmission block, where $L=N+P-1$.

The underlying assumptions in the system we consider are given below. It is assumed that perfect frame synchronization can be achieved, and adequate guard periods are added between consecutive transmission blocks so that
there is no inter-block interference. In addition, both the
transmitter and the receiver know nothing about the channel
coefficients $\hb$ except the multipath parameter $P$.
Notably, the training sequence does not have to be placed at the beginning
of $\bb$, but can be distributed over the entire transmission
block. It however has been shown that placing the training
sequence at the beginning of $\bb$, together with
$\Bb_P^\HH\Bb_P=T\Ib$, can minimize the variance of estimation
error \cite{CWS}. This justifies the model in \eqref{channel_model}, where
$\Bb_P$ is placed ahead of $\Bb_D$.
The condition $\Bb_P^\HH\Bb_P=T\Ib$ is accordingly assumed following \cite{CWS}.

\bigskip

\section{Criterion and algorithm of the proposed two-phase method}

Based on the system model in \eqref{channel_model}, we can divide
the received signal $\yb$ into two parts:
$$\begin{cases}
\yb_P=\Bb_P\hb+\nb_P\\
\yb_D=\Bb_D\hb+\nb_D\end{cases}$$
where $\yb_P$ and $\yb_D$ are defined via
$\yb^\HH=\begin{bmatrix}\yb_P^\HH& \yb_D^\HH\end{bmatrix}$,
and $\nb_P$ and $\nb_D$ are similarly defined.
Under the reasonable premise that $T\geq P$, the least square estimate of $\hb$, given $\Bb_P$ and $\yb_P$, is equal to
$$\hat{\hb}=(\Bb_P^\HH\Bb_P)^{-1}\Bb_P^\HH\yb_P.$$
Then the ML decoding
criterion for a receiver with only partial CSI is given by \cite{CWS}:
\begin{eqnarray}
\hat{\bb}_{\rm ML}&=&\arg\max_{\Bb_D}\Pr(\yb_D|\Bb_D,\hb=\hat{\hb})\nonumber\\
&=&\arg\min_{\Bb_D}\bigg(\|\yb_D-\Bb_D\hat{\hb}\|^2\nonumber\\
&&\quad-(\yb_D-\Bb_D\hat{\hb})^\HH\Qb_{B}(\yb_D-\Bb_D\hat{\hb})\nonumber\\
&&\quad+\sigma_n^2\log\det\left(\Ib+(\Bb_P^\HH\Bb_P)^{-1}\Bb_D^\HH\Bb_D\right)\bigg),\label{metric}
\end{eqnarray}
where
$$\Qb_{B}=\Bb_D\left(\Bb_D^\HH\Bb_D+\Bb_P^\HH\Bb_P\right)^{-1}\Bb_D^\HH.$$
At medium to high SNRs, the last term in \eqref{metric}
becomes negligible when it is compared with the first two terms; hence,
a near-ML decoding criterion can be yielded as follows:
\begin{eqnarray}
\hat{\bb}_{\text{near-ML}}&=&\arg\min_{\Bb_D}\left\{\|\yb_D-\Bb_D\hat{\hb}\|^2\right.\nonumber\\
&&\left.-\left(\yb_D-\Bb_D\hat{\hb}\right)^\HH\Qb_{B}\left(\yb_D-\Bb_D\hat{\hb}\right)\right\}.\label{near_ML_metric}
\end{eqnarray}
It is noted that the criteria for both $\hat{\bb}_{\rm ML}$ and
$\hat{\bb}_{\text{Near-ML}}$ contain the Euclidean distance
$\|\yb_D-\Bb_D\hat{\hb}\|^2$ as their first term, which can be
easily decomposed into \emph{finite-state} recursive expression
that readily suits the need of the VA. However, the remaining
terms in \eqref{metric} and \eqref{near_ML_metric} do not have
\emph{finite-state} recursive expressions, so the VA cannot be
applied to obtain either $\hat{\bb}_{\rm ML}$ or
$\hat{\bb}_{\text{Near-ML}}$.

At this background, we propose a two-phase method to perform
sequence estimation for a partially coherent receiver. The first phase
is exactly the MLSE using the Euclidean distance $\phi_{L-T}\equiv\|\yb_D-\Bb_D\hat{\hb}\|^2$ in recursive form, i.e.,
\begin{eqnarray}
\phi_{t}=\phi_{t-1}+|y_{t+T}-\ub_t^{\HH}\hat{\hb}|^2,\label{forward_metric_without_state}
\end{eqnarray}
where ``$\equiv$" denotes that the two sides are
equivalent metrics in decoding, and $y_{t+T}$ and $\ub_t^\HH$ are
respectively the $(t+T)$th component of $\yb$ and the $t$th row of
$\Bb_D$. In order to apply the recursive metric in
\eqref{forward_metric_without_state} on a VA trellis,
we reformulate the accumulated metric $\phi_{t}$ as a function
of the \emph{trellis state} $i$ as follows:
\begin{eqnarray}
\phi_{t}(i)=\min_{1\leq j \leq
2^{M(P-1)}}\left\{\phi_{t-1}(j)+|c_t(j,i)|^2
\right\},\label{forward_metric}
\end{eqnarray}
where $t$ and $i$ are respectively ranged from $1$ to $L-T$ and from $1$ to $2^{M(P-1)}$,
\begin{equation}
c_{t}(j,i)=y_{t+T}-\ub_{t}(j,i)^{\HH}\hat{\hb},\label{ct}
\end{equation}
and $\ub_t(j,i)=[b_{t+T-P+1}(j,i),\cdots,b_{t+T}(j,i)]$
denotes the signals corresponding
to the trellis branch from state $j$ at time $t-1$ to state $i$ at time $t$.
Meanwhile, two variables will be calculated during
the execution of the first phase so that they can be used in the
second phase, which are:
$$
\begin{cases}
\eta_{t}^{(\ell)}(i)=\eta_{t-1}^{(\ell)}(j)+c_t(j,i)\cdot
b_{t+T+\ell}(j,i)\\
\rho_{t}^{(\ell)}(i)=\rho_{t-1}^{(\ell)}(j)+\left(b_{t+T}(j,i)\right)^{*}\cdot
b_{t+T-\ell}(j,i)
\end{cases}
$$
where in the above two formulas,  $j$ is the minimizer of \eqref{forward_metric}, and $t$ and $\ell$ are ranged from $1$ to $L-T$ and from $0$ to $P-1$, respectively.

In the second phase, a backward VA is performed. Since the
simulations in \cite{CWS} show that \eqref{metric} and
\eqref{near_ML_metric} yield almost the same performance, we adopt
the criterion in \eqref{near_ML_metric} to save the computational
complexity.
We then reexpress the criterion in
\eqref{near_ML_metric} into an indirect backward recursive form:
\begin{equation}
\Sigma_{t}(j,i)=\varphi_{t+1}(j)-\lambda_{t}(j,i)+\phi_{t}(i)+|c_{t+1}(i,j)|^2,\label{backward_metric}
\end{equation}
where $j$ and $i$ are respectively the previous and current
states that define the concerned branch in the backward trellis,
$c_{t+1}(i,j)$ is defined in \eqref{ct},
and except that $\phi_t(i)$ is from the first phase, the other two terms (i.e., $\varphi_{t+1}(j)$ and $\lambda_{t}(j,i)$) are backward-recursively computed as follows.
By letting
\begin{eqnarray}
\xi_{t}(i)=\arg\min_{1\leq j \leq 2^{M(P-1)}}\Sigma_{t}(j,i),\label{backward_detection}
\end{eqnarray}
we have
$$
\varphi_{t}(j)=\varphi_{t+1}(\xi_{t}(j))+|c_{t+1}(j,\xi_{t}(j))|^2,$$
and
$$\lambda_{t}(j,i)=\sum_{u=0}^{P-1}\sum_{v=0}^{P-1}\delta_{u,v}(j,i)\cdot
\left({r_t^{(u)}(j,i)}\right)^{*} r_t^{(v)}(j,i),$$
where $\delta_{u,v}(j,i)$ is the entry at the $u$th row and the $v$th
column of matrix $\Db(j,i)^{-1}$,

\bigskip

\noindent$
\Db(j,i)=\Bb_{P}^{\HH}\Bb_{P}+$

\bigskip

\noindent$
\begin{bmatrix}
\psi_{t}^{(0)}(j,i)-P+1 & \psi_{t}^{(1)}(j,i) & \!\!\!\cdots \!\!\!& \psi_{t}^{(P-1)}(j,i)\\
{\psi_{t}^{(1)}(j,i)}^{*} & \psi_{t}^{(0)}(j,i)-P+2 & \!\!\!\cdots \!\!\!&
\psi_{t}^{(P-2)}(j,i)\\
\vdots & \vdots & \!\!\!\ddots \!\!\!&
\vdots\\
{\psi_{t}^{(P-1)}(j,i)}^{*} & {\psi_{t}^{(P-2)}(j,i)}^{*} & \!\!\!\cdots\!\!\!
& \psi_{t}^{(0)}(j,i)
\end{bmatrix},
$

$$
r_t^{(\ell)}(j,i)=\eta_{t}^{(\ell)}(i)+\zeta_{t+1}^{(\ell)}(j)+c_{t+1}(i,j)\cdot
b_{t+1+T-\ell}(i,j),
$$
$$
\zeta_{t}^{(\ell)}(j)=\zeta_{t+1}^{(\ell)}(\xi_{t}(j))+c_{t+1}(j,\xi_{t}(j))\times
b_{t+1+T-\ell}(j,\xi_{t}(j)),
$$
$$
\psi_{t}^{(\ell)}(j,i)=\sigma_{t+1}^{(\ell)}(j)+\left(b_{t+1+T}(i,j)\right)^{*}\cdot
b_{t+1+T-\ell}(i,j)+\rho_{t}^{(\ell)}(i),
$$
and
$$
\sigma_{t}^{(\ell)}(j)=\sigma_{t+1}^{(\ell)}(\xi_{t}(j))+\left(b_{t+1+T}(j,\xi_{t}(j))\right)^{*}\cdot
b_{t+1+T-\ell}(j,\xi_{t}(j)).
$$

We end this section by summarizing our proposed two-phase method
in an algorithmic form.

\bigskip

\noindent \emph{\textbf{The First Phase (Forward VA):}}
\begin{list}{Step~\arabic{step}.}
    {\usecounter{step}
    \setlength{\labelwidth}{0.2cm}
    \setlength{\leftmargin}{0.5cm}\slshape}
\item [Step 1-1. Initialization:] ~

For $1\leq\ell\leq{P-1}$ and for
$1\leq i\leq 2^{M(P-1)}$, initialize $\eta_{0}^{(\ell)}(i)=0$ and
$\rho_{0}^{(\ell)}(i)=0$ . Let $\phi_0(1)=0$ and
$\phi_0(i)=\infty$ for $2\leq i\leq 2^{M(P-1)}$.

\item [Step 1-2. Recursion (From $t=1$ to $t=L-T$):] ~

For
$1\leq i \leq 2^{M(P-1)}$ and
for $1\leq \ell <P$, compute
$$
\begin{cases}
\phi_t(i)=\min_{1\leq j \leq
2^{M(P-1)}}\left(\phi_{t-1}(j)+|c_t(j,i)|^2\right),\\
\xi_t(i)=\arg\min_{1\leq j \leq
2^{M(P-1)}}\left(\phi_{t-1}(j)+|c_t(j,i)|^2\right).\\
\end{cases}
$$
Update
$$
\begin{cases}
\eta_t^{(\ell)}(i)=\eta_{t-1}^{(\ell)}(\xi_t(i))+c_{t}(\xi_t(i),i)\cdot
b_{t+T-\ell}(\xi_t(i),i),\\
\rho_t^{(\ell)}(i)=\rho_{t-1}^{(\ell)}(\xi_t(i))+{\left(b_{t+T}(\xi_t(i),i)\right)}^{*}
b_{t+T-\ell}(\xi_t(i),i),
\end{cases}
$$ where
$$c_t(j,i)=y_{T+t}-\ub(j,i)^{\HH}\hat{\hb}$$
and $\ub(j,i)=[b_{t+T-P+1}(j,i),\cdots,b_{t+T}(j,i)]$ consists
of  $P$ symbols corresponding to the trellis branch between
state $j$ at time $t-1$ and state $i$ at time $t$.
\end{list}

\bigskip

\newpage

\noindent\emph{\textbf{The Second Phase (Backward VA):}}
\begin{list}{Step~\arabic{step}.}
    {\usecounter{step}
    \setlength{\labelwidth}{0.2cm}
    \setlength{\leftmargin}{0.5cm}\slshape}
\item [Step 2-1. Initialization:] ~

For $1\leq\ell\leq{P-1}$ and for
$1\leq i\leq 2^{M(P-1)}$, initialize $\zeta_{L-T+1}^{(\ell)}(i)=0$
and $\sigma_{L-T+1}^{(\ell)}(i)=0$. Let $\varphi_{L-T+1}(1)=0$ and
$\varphi_{L-T+1}(i)=\infty$ for $2\leq i\leq 2^{M(P-1)}$.

\item [Step 2-2. Recursion (From $t=L-T$ down to $t=1$):] ~

For $1\leq
i \leq 2^{M(P-1)}$, compute
$$
\begin{cases}
\Sigma_t(j,i)=\left(\varphi_{t+1}(j)+|c_{t+1}(i,j)|^2\right)+\phi_{t}(i)-\lambda_{t}(j,i),\\
\xi_t(i)=\arg\min_{1\leq j \leq 2^{M(P-1)}}\Sigma_t(j,i),
\end{cases}
$$
where the terms involved in the above computations have been introduced previously.

Update
$$
\begin{cases}
\varphi_{t}(i)=\varphi_{t+1}(\xi_t(i))+|c_{t+1}(i,\xi_t(i))|^2\\
\zeta_{t}^{(\ell)}(i)=\zeta_{t+1}^{(\ell)}(\xi_t(i))+c_{t+1}(i,\xi_t(i))
b_{t+1+T-\ell}(i,\xi_t(i))\\
\sigma_{t}^{(\ell)}(i)=\sigma_{t+1}^{(\ell)}(\xi_t(i))\\
~~~~~~~~~~~~~~~~+\left(b_{t+1+T}(i,\xi_t(i))\right)^{*}b_{t+1+T-\ell}(i,\xi_t(i))
\end{cases}
$$

\item [Step 2-3. Trace Back:] ~

Output the best state sequence
$[1,$ $s_1,$ $\cdots,$ $s_{L-T},$ $1]$, where $s_t=\xi_t(s_{t-1})$, and its
corresponding decision symbol sequence.
\end{list}

\bigskip

\section{Complexity Analysis}

The computational complexity of the proposed algorithm consists of
the forward VA complexity $C_{\text{F}}$ and backward VA
complexity $C_{\text{B}}$. Since both the forward VA and backward
VA operate on a trellis having $2^{M(P-1)}(N-T)$ states and there
are $2^M$ branch metric calculations for each state, these two
complexities can be expressed as
$$C_{\text{F}}=N_{\text{F}}\cdot 2^M\cdot 2^{M(P-1)} (N-T)$$ and
$$C_{\text{B}}=N_{\text{B}}\cdot 2^M\cdot 2^{M(P-1)}(N-T),$$ where
$N_{\text{F}}$ and $N_{\text{B}}$ are the branch metric
computational complexities in forward VA and backward VA,
respectively.

By convention, the complex multiplications dominate the branch
metric computational complexity; therefore, $N_{\text{F}}$ and
$N_{\text{B}}$ can be approximated by the number of complex
multiplications required in forward VA and backward VA,
respectively. As a result, in forward VA, there is a $P$-tag
filter and two additional complex multiplications for each branch;
so, we set $N_{\text{F}}=P+2$. In backward VA, each branch metric
calculation needs a $P$-tag filter for the calculation of
$\varphi_t(\cdot)$, $2 P^2$ complex multiplications for
$\lambda_t(\cdot,\cdot)$, $P^3$ complex multiplications for
$\delta_{u,v}(\cdot,\cdot)$ and $4P$ complex multiplications for
the remaining variables. We then obtain
$N_{\text{B}}=5P+2P^2+P^3$. The total computational complexity is
accordingly given by
 \begin{eqnarray}
\lefteqn{C_{\text{F}}+C_{\text{B}}=(N_{\text{F}}+N_{\text{B}})\cdot 2^{M P}(N-T)}\nonumber\\
&=(2+6P+2P^2+P^3) \cdot 2^{M P} (N-T).
 \end{eqnarray}
 The complexity is considerably more than the complexity of conventional MLSE, which is $P\cdot 2^{M P} (N-T)$. However, it is much smaller than the complexity of the optimal exhaustive checking decoder, which is
 \begin{equation}
 (N_{\text{F}}+N_{\text{B}})\cdot 2^{MN+1}
 \end{equation}

We remark at the end that because the complexity of on-line computations of $\delta_{u,v}(\cdot,\cdot)$
is high for a large $P$, the proposed two-phase method may be more suitable for channels with small $P$, or for a system with
pre-filters at the receiver to reduce the tap number of channels
\cite{Gerstacker,Falconer}.

\bigskip

\section{Simulations}

For simplicity, only BPSK modulation is considered in simulations; thus $M=1$.
The channel coefficients $\hb$ are zero-mean complex-Gaussian
distributed with $E[\hb\hb^\HH]=(1/P)\Ib$ and $P=2$.
By the system model introduced in Section \ref{systemmodel},
the signal-to-noise power ratio per information bit
is given by
\begin{eqnarray*}
\lefteqn{\frac{E_\text{b}}{N_0}\text{ (dB)}=10\log_{10}\frac{\tr\left(E[\hb\hb^\HH]\right)}{\sigma_n^2}}\\
&=&10\log_{10}\frac{E[\hb^\HH\hb]}{\sigma_n^2}=10\log_{10}\frac{1}{\sigma_n^2}.
\end{eqnarray*}

\begin{figure}
    \begin{center}
    \includegraphics[width=0.49\textwidth,height=0.32\textheight]{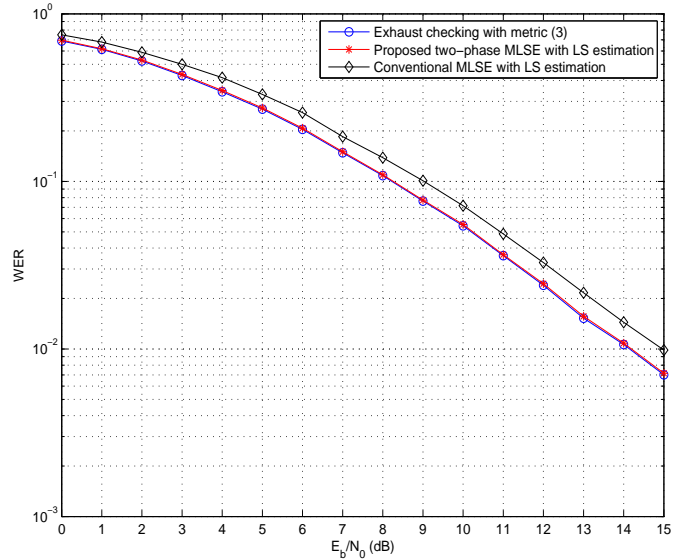}
    \caption{Word error rates (WERs) of three MLSE schemes in block fading channels for a data burst of length $15$, in which 5 of them are training sequence.}
    \label{f0}
    \end{center}
\end{figure}

We first examine our proposed two-phase method
using a data sequence of length $N=15$, in which
$5$ of them are training sequence and are equal to $[-1,-1,-1,1,-1]$.
Figure~\ref{f0} then shows that the word error rate (WER) of our proposed
two-phase method is almost the same as that of
the exhaustive checking scheme using criterion \eqref{near_ML_metric}.
This figure also indicates that our proposed two-phase method
outperforms the conventional MLSE by about $0.8$ dB.
All three schemes estimate channel
coefficients via a least square (LS) estimator.
This result confirms that our proposed two-phase MLSE
(designed based on criterion \eqref{near_ML_metric} for complexity saving) can achieve
the optimal performance of exhaustive checking when the length of the training sequence is moderately large (for example, 1/3) in comparison with the entire block size.

\begin{figure}
    \begin{center}
    \includegraphics[width=0.49\textwidth,height=0.32\textheight]{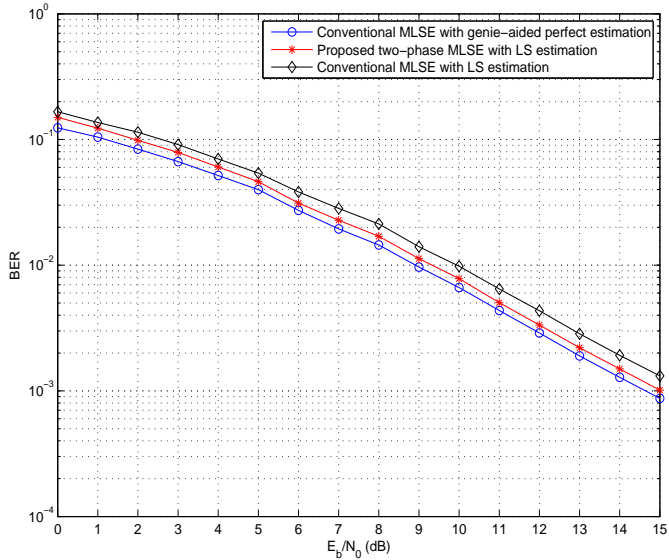}
    \caption{Bit error rates (BERs) of three MLSE schemes in block fading channels for a data burst of length $70$, in which 10 of them are training sequence.}
    \label{f1}
    \end{center}
\end{figure}

Next, we consider a longer block of length $N=70$, in which only
10 of them are training sequence and are equal to
$[1,1,1,1,1,-1,1,-1,1,-1]$. Note that the training sequence
consumes around $10/70=0.14$ of the bandwidth.\footnote{ This
number is smaller than what is considered in a GSM data burst,
where a $148$-bit normal burst contains a $26$-bit training
sequence. } Figure \ref{f1} then shows that the proposed two-phase
method still maintains a $0.7$ dB advantage in comparison with the
conventional MLSE with LS estimation. Because at
this block length, the exhaustive checking method is no longer
feasible, we provide the performance of the conventional MLSE with
perfect CSI in this figure as a reference genie-aided performance
lower bound.

\begin{figure}
    \begin{center}
    \includegraphics[width=0.49\textwidth,height=0.32\textheight]{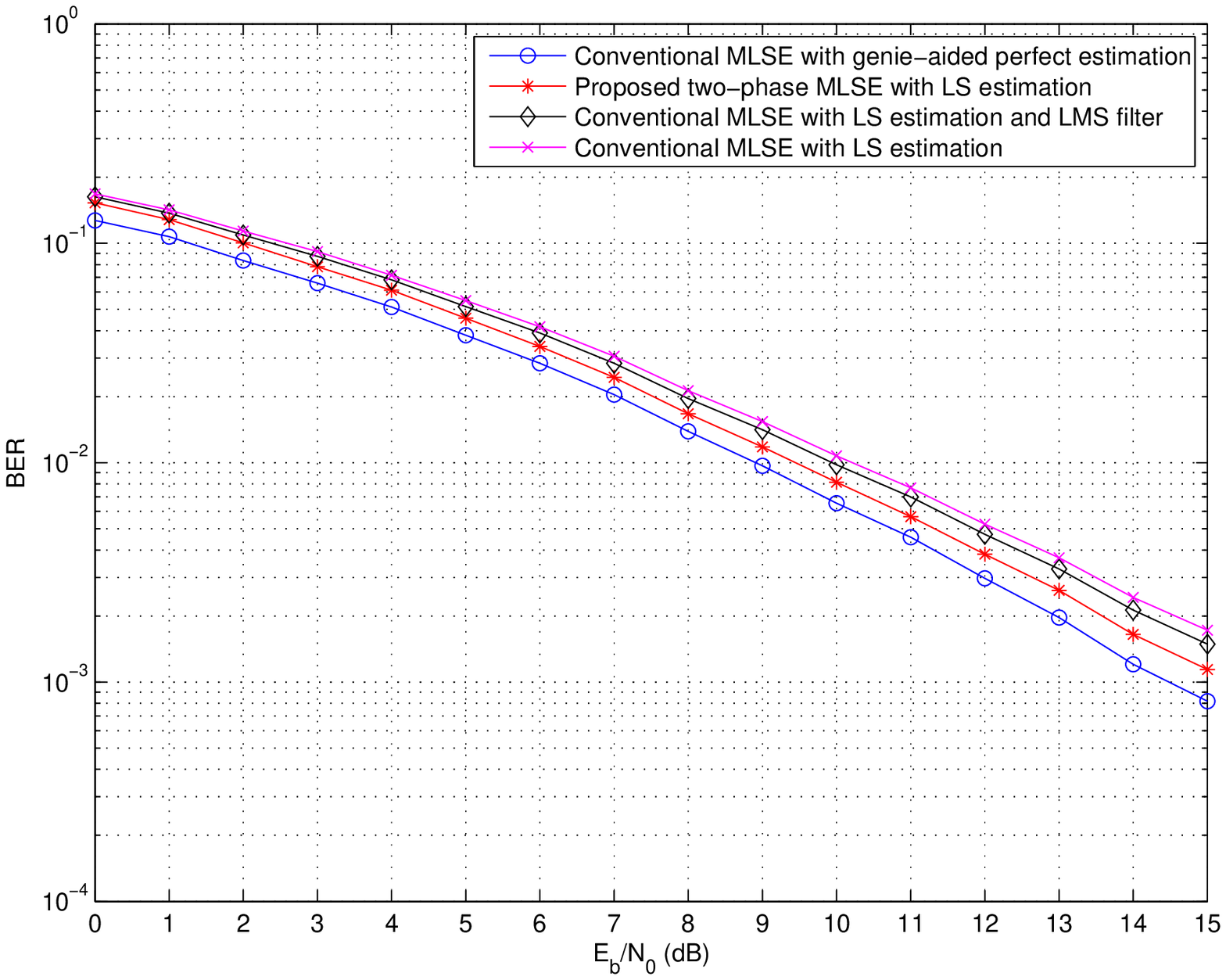}
    \caption{Bit error rates (BERs) of three MLSE schemes in Gauss-Markov  channels with $\alpha=0.9999$ for a data burst of length $70$, in which 10 of them are training sequence.}
    \label{f5}
    \end{center}
\end{figure}

\begin{figure}
    \begin{center}
    \includegraphics[width=0.49\textwidth,height=0.32\textheight]{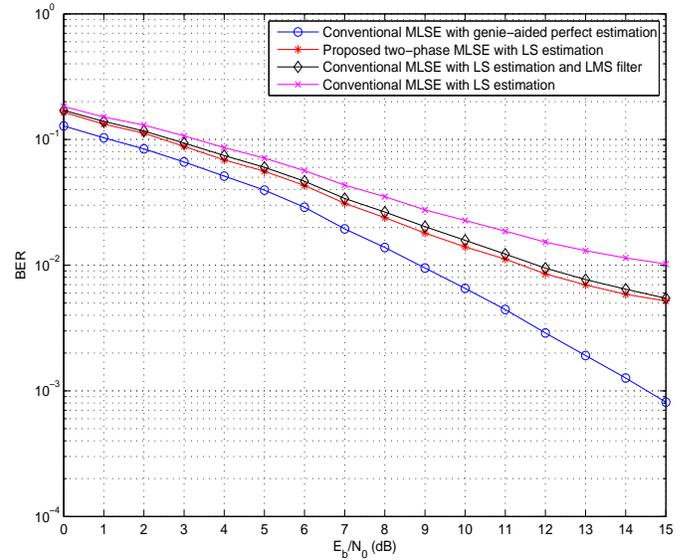}
    \caption{Bit error rates (BERs) of three MLSE schemes in Gauss-Markov  channels with $\alpha=0.999$ for a data burst of length $70$, in which 10 of them are training sequence.}
    \label{f6}
    \end{center}
\end{figure}

In Figs.~\ref{f5} and \ref{f6}, we examine our proposed
two-phase method over the Gauss-Markov fading channel
\cite{Misra,Zhang}. In this channel, the channel coefficients that
are fixed within a data burst period are varying according to
\begin{equation}
\label{GM}
\hb_t=\alpha\cdot \hb_{t-1}+\sqrt{1-\alpha^2}\vb_t
\end{equation}
for $1\leq t\leq L$, where $\hb_0$ and $\vb_t$ are independent to each other and are zero-mean Gaussian random vectors with covariance matrix $(1/P)\Ib$. By \eqref{GM},
it can be easily verified that
the SNR per information bit remains:
$$\frac{E_{\text{b}}}{N_0}(dB)=10\log_{10}\frac{E[\hb_t^\HH\hb_t]}{\sigma_n^2}=10\log_{10}\frac{1}{\sigma_n^2}.$$
The data format tested in Figs.~\ref{f5} and \ref{f6} is the same
as that used in Fig.~\ref{f1}. An additional scheme is added in
comparison with our proposed two-phase method, which is the MLSE
with an adaptive least mean square (LMS) filter
\cite{Chiu,Proakis}. This filter has been proved to be effective
in tracking the time-varying nature of time-varying channels. Under
the assumption that the receiver can perfectly estimate the value
of $\alpha$, the step size of the LMS filter used in our
simulations is set to be $\sqrt{1-\alpha^2}/2.$

Figure \ref{f5} then shows that for $\alpha=0.9999$, our two-phase
method outperforms the other two equalization schemes. The
simulation result under $\alpha=0.999$ also indicates similar
performance gain of our two-phase method over the other two
equalization schemes except that a performance floor appears at
high SNR. We again provide the performances of the conventional
MLSE with perfect CSI in these two figures as reference
genie-aided performance lower bounds.

\bigskip

\section{Conclusion}

After establishing the recursive expression of ML criterion for partially
coherent receiver, we propose a two-phase
MLSE algorithm in this paper. Simulation results show that our method
outperforms the conventional MLSE in both quasi-static block fading channels
and time-varying Gauss-Markov channels.
A possible future work could be to modify our algorithm to provide soft-outputs so that it can iteratively co-work with an outer coding scheme.

\bigskip

\section{Acknowledgment}

The first author would like to thank to the financial support from the
Postdoctoral Research Abroad Program sponsored by the National
Science Council of Taiwan under the Grant NSC
101-2917-I-564-006.

\end{document}